\documentclass{cupconf}
\usepackage{psfig}
\title[Hubble's View of Transiting Planets]{Hubble's View of Transiting Planets}
\author[D. Charbonneau]{D\ls A\ls V\ls I\ls D\ns 
C\ls H\ls A\ls R\ls B\ls O\ls N\ls N\ls E\ls A\ls U\ls}
\affiliation{Harvard-Smithsonian Center for Astrophysics,
60 Garden Street, Cambridge, MA 02138 USA}
\begin{document}
\maketitle
\begin{abstract}
The \emph{Hubble Space Telescope} is uniquely able to study
planets that are observed to transit their parent
stars.  The extremely stable platform afforded by
an orbiting spacecraft, free from the contaminating
effects of the Earth's atmosphere, enables \emph{HST} to
conduct ultra-high precision photometry and spectroscopy 
of known transiting extrasolar planet systems.  Among \emph{HST}'s
list of successful observations of the first such system, 
HD~209458, are (1) the first detection of the atmosphere of 
an extrasolar planet, (2) the determination
that gas is escaping from the planet, and (3) a search for 
Earth-sized satellites and circumplanetary rings.
Numerous wide-field, ground-based transit surveys
are poised to uncover a gaggle of new worlds for
which \emph{HST} may undertake similar studies, such as the newly-discovered
planet TrES-1.  With regard
to the future of \emph{Hubble}, it must be noted that
it is the only observatory in existence capable of 
confirming transits of Earth-like planets that
may be detected by NASA's \emph{Kepler} mission.  \emph{Kepler} could
reveal Earth-like transits by the year 2010, but
without a servicing mission it is very unlikely that
\emph{HST} would still be in operation.
\end{abstract}
\firstsection
\section{Introduction}
When both the photometric transits and radial velocity variations
due to an extrasolar planet are observed, we are granted access
to key quantities of the object that Doppler monitoring alone cannot
provide.  In particular, precise measurements
of the planetary mass and radius allow us to calculate the average
density, and infer a composition.  Such estimates enable a meaningful
evaluation of structural models of these objects, including whether
or not they possess a core of rocky material.  These inferences, in turn,
enable direct tests of competing scenarios of planet formation and evolution.
Moreover, the transiting configuration permits numerous interesting follow-up 
studies, such as searches for planetary satellites and circumplanetary rings,
and studies of the planetary atmosphere.

In this review presented at the Space Telescope Science Institute's 2004 May 
Symposium \emph{``From Planets to Cosmology: Essential Science in Hubble's Final Years''},
I discuss the status of work in the field with a focus on the 
key contributions, both past and near-future, enabled by the \emph{Hubble Space Telescope}.  
I begin by reviewing the properties of the 
transiting planets discovered to date, as well as the the numerous
ground-based efforts to detect more of these objects.  I then
consider the various follow-up studies of these gas-giant planets 
that are enabled by \emph{HST}, as well as \emph{HST}'s potentially
unique role in following-up Earth-sized objects detected by NASA's
\emph{Kepler} mission.  I finish by discussing two \emph{HST}-based searches for 
transiting gas-giant planets.  Throughout this contribution, I restrict my attention 
to studies of \emph{transiting} planets; for an introduction to 
the broader range of \emph{HST}-based studies of extrasolar planets, see Charbonneau (2004).

\section{The Current Sample of Transiting Planets}
At the time of writing, there are 6 extrasolar planets known to
transit the disks of their parent stars:
HD~209458b (Charbonneau et al. 2000, Henry et al. 2000) was initially
identified through Doppler monitoring, TrES-1 (Alonso et al. 2004)
was discovered by the small-aperture, wide-field Trans-Atlantic Exoplanet Survey
(TrES) network, and the four others were found by Doppler follow-up of a 
list of more than 100 candidates identified by the OGLE team
(Udalski 2002a, 2002b, 2002c, 2003): OGLE-TR-56b (Konacki et al. 2003), 
OGLE-TR-111b (Pont et al. 2004), OGLE-TR-113b (Bouchy et al. 2004; Konacki et 
al. 2004), and OGLE-TR-132b (Bouchy et al. 2004).

As a result of the recent flurry of discoveries, we may construct a mass-radius plot 
for transiting extrasolar planets, shown in Figure~1.  From this plot,
it is clear that all but one of the planets have radii that are consistent with
a value that is only modestly inflated (typically $<$~15\%) over the Jupiter radius.  This
is in keeping with the predictions of models which include the effects
of stellar insolation (Guillot et al. 1996; Burrows et al. 2000; Baraffe et al. 2003;
Bodenheimer et al. 2003; Chabrier et al. 2004).  There is a growing consensus among the 
theorists that the large radius of HD~209458b, $1.35 \pm 0.06 \ R_{\rm Jup}$,
cannot be explained by stellar insolation effects alone, and indeed Figure~1
points to its anomalous nature.  Guillot \& Showman (2002) and Showman \& Guillot (2002)
considered a model in which a modest fraction of the incident stellar radiation
in converted into mechanical energy (winds), which could fill the energy
decrement and hence explain the large radius.  It is not clear, however,
why this mechanism would not work as efficiently for the other objects.
An alternative hypothesis is that the orbital eccentricity of HD~209458b 
is continually pumped by the presence of a more distant and, as of yet,
undetected planet; the damping of the eccentricity in turn provides the
energy source needed to explain the large radius (Bodenheimer et al. 2003).
Fortunately, it should be possible to test this model observationally, 
initially through the offset in the timing of the
secondary eclipse (indicating a non-zero orbital eccentricity; 
Charbonneau 2003b), and subsequently through the detection of the 
second planet through precise radial velocity measurements.
Solving the mystery of HD~209458b will be an engaging challenge with a resolution 
likely in the next couple years.

HD~209458b and TrES-1 are very similar in mass and equilibrium
temperature, yet have dramatically different radii.
As a result, we must abandon the simple expectation 
that the radius of an extrasolar gas-giant planet will be determined
solely by its mass and degree of stellar insolation.  Rather,
we now know that the radii can be altered dramatically
by additional processes.  Identifying these mechanisms and
understanding their implications for the structure of these
planets will be a rewarding task for the near future.  

As discussed in Section 4, only relatively nearby systems will be bright 
enough for the majority of follow-up observations with \emph{HST} that
have been completed for HD~209458b.  Indeed, of the
6 known systems, only HD~209458b and TrES-1 satisfy the brightness
criterion.  Since we would like to increase the number of
similarly-accessible planets, I consider next the current ground-based
surveys for these objects.

\begin{figure}
\centerline{\psfig{figure=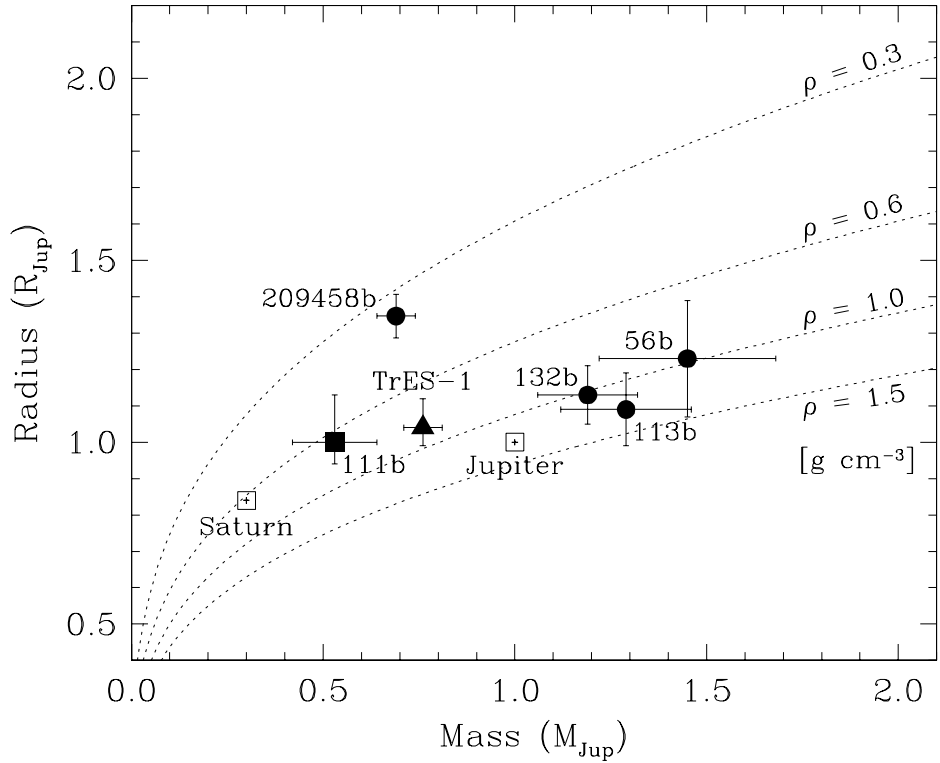,width=5.6in}}
\caption{\emph{Reprinted with permission from Sozzetti et al. (2004).} Radius versus
mass for all known transiting planets.  The values and uncertainties
are taken from Brown et al. (2001) for HD~209458b, Sozzetti et al. (2004)
for TrES-1, Torres et al. (2004a) for OGLE-TR-56b, Pont et al. (2004) for 
OGLE-TR-111b, Bouchy et al. (2004) for OGLE-TR-113b, and Moutou et al. (2004) 
for OGLE-TR-132b.  Lines of constant density are also indicated.}
\end{figure}

\section{Ground-Based Surveys for New Targets}
Of the roughly 20 ground-based transit surveys (Horne 2003, 
Charbonneau 2003a)\footnote{\texttt{http://star-www.st-and.ac.uk/$\sim$kdh1/transits/table.html}}
that are either in operation or planned for the near future,
only those that survey bright ($V<13$) stars will find planets
that may be pursued with the \emph{HST}-based techniques described in the
following section. Among these surveys are SuperWASP (Street et al. 2003), 
PASS (Deeg et al. 2004), HATnet (Bakos et al. 2004), Vulcan (Borucki et al. 2001), 
KELT (Pepper et al. 2003), and TrES (e.g. Alonso et al. 2004).  All of these
surveys use fast optics (typically consisting of a commercially-available
camera lens) and a correspondingly coarse pixel scale (typically $10-20$~arcsec~/~pixel)
to monitor thousands (and sometimes tens of thousands)
of field stars.  The dominant challenge facing such searches is no longer
that of obtaining the requisite photometric precision (better than 1\%)
and sufficient phase coverage (typically 300 hours) on a sufficient
number of stars ($>$~5000), but rather that of identifying the astrophysical
false positives (Latham 2003; Charbonneau et al. 2004; 
Mandushev et al. 2004; Torres et al. 2004b), which are 
predicted to occur with a frequency
of roughly 10 times that of true planetary systems (Brown 2003).
These impostors each consist of a stellar system 
containing an eclipsing binary that precisely mimics the single-band light
curve of a gas-giant planet transiting a Sun-like star.  Fortunately, the stars 
surveyed by these projects are bright, which has several ramifications.
First, the targets typically have well-determined 
2MASS\footnote{\texttt{http://www.ipac.caltech.edu/2mass/releases/allsky/}} colors 
and USNO-B proper
motions (Monet et al. 2003)\footnote{\texttt{http://www.nofs.navy.mil/projects/pmm/catalogs.html}}, 
so that the likelihood that a given target is a nearby late-type
dwarf can be readily evaluated.  Second, the stars are easily accessible to spectrographs mounted 
on modest-aperture telescopes, where observing time is often easily
available:  Initial spectroscopic monitoring with a lower Doppler
precision ($\sim 1 \ \rm{km \ s^{-1}}$; Latham 2003) is one effective way to
reject the majority of such contaminants before gathering precise ($\sim 10 \ \rm{m \ s^{-1}}$)
radial velocity observations at much larger observatories where observing time 
is often heavily over-subscribed.

\subsection{The TrES Network}
The Trans-Atlantic Exoplanet Survey (TrES) consists of 3 telescopes: 
Sleuth\footnote{\texttt{http://astro.caltech.edu/$\sim$ftod/tres/sleuth.html}} 
(Charbonneau 2003a; O'Donovan et al. 2004),
STARE\footnote{\texttt{http://www.hao.ucar.edu/public/research/stare/stare.html}} 
(Brown \& Charbonneau 2000), and PSST (Dunham et al. 2004).  Each instrument
was developed independently but with the intent that all three could monitor the
same field of view with very similar capabilities.  The Sleuth telescope consists
of an $f$/2.8 commercial camera lens with an aperture of 10~cm.  It images a $5.7^{\circ} \times
5.7^{\circ}$ patch of sky onto a 2~k~$\times$~2~k thinned CCD, with a resulting pixel
scale of 9.9~arcsec.  Sleuth gathers data in the SDSS~$r$ filter (the other systems
use Bessell~$R$), but has a filter wheel with additional filters so that accurate
colors of each target may be obtained.  The system is located in an automated clamshell
enclosure at Palomar Observatory, and each night's observing sequence is handled
in a completely automated fashion by a local workstation running Linux.  Due to its
automated operation, Sleuth achieves a very high efficiency, gathering data
on nearly every night weather permits.  Calibration
images are also obtained nightly.  Each morning the entire sequence
of images is compressed and transferred to the analysis workstation at Caltech.
The other systems in the TrES network operate in a roughly similar fashion.

Over the past year, we have switched our analysis pipeline from one based 
on weighted-aperture photometry (i.e. Brown \& Charbonneau 2000) to 
that of image subtraction (i.e. Allard 2000).  This change has resulted in much better 
performance, and produces time series with an rms very near the
theoretical expectation.  As a result, we have significantly increased 
the number of stars that we survey with the requisite photometric precision
to roughly $10,000 - 15,000$ per field (near the galactic plane).  

In order to handle the large number of astrophysical false positives
identified by the TrES survey, we have recently assembled a new
telescope named Sherlock (Charbonneau et al. 2004; Kotredes et al. 2004) that will be dedicated
to such follow-up work.  Sherlock is located in the same enclosure
as Sleuth at Palomar Observatory.  It is based on a commercially available $f/6.3$
Schmidt-Cassegrain telescope with an aperture of 25~cm.  It images a 
$0.5^{\circ}~\times~0.5^{\circ}$ field-of-view onto a 1~k~$\times$~1~k
thinned CCD, and is equipped with a filter wheel housing a selection
of relatively narrow-band interference filters.  Each night, we 
identify any active candidates that are predicted to present eclipses,
and perform high-cadence multi-color photometry of the highest-priority
target.  Sherlock is able to distinguish several forms of false positives
for two reasons:  First, the increased angular resolution (1.7~arcsec / pixel)
compared to the TrES instruments (10~arcsec / pixel) resolves most
stellar blends that occur due to projection along the line-of-sight;
in such a scenario, the light curve of the isolated source as seen by Sherlock typically
presents a significantly deeper eclipse, which is no longer consistent with
a transiting planet.  Second, stellar blends (physically associated or not)
in which the blending and occulted stars are of significantly different
temperature will present eclipse depths that vary in color in excess
of the small effects due to limb-darkening.  The follow-up photometry
provided by Sherlock is complementary to spectroscopic work similarly
aimed at identifying the impostors (i.e. Latham 2003).  Notably, we plan to operate Sherlock
in a fully-automated manner, and the small amount of labor compared to
the spectroscopic follow-up makes this method an appealing one by which
to reject the bulk of such astrophysical false positives.

\subsection{The New World of TrES-1}
Alonso et al. (2004) present the detection of the planet TrES-1,
a transiting hot Jupiter with an orbital period of 3.03~days.
This work, combined with a detailed analysis
of the stellar spectrum by Sozzetti et al. (2004), permits
a precise estimation of the planetary mass $M_p = 0.76\pm 0.05 \ M_{\rm Jup}$ 
and radius $R_p = 1.04^{+0.08}_{-0.05} \ R_{\rm Jup}$.
As noted above, the 3-$\sigma$ discrepancy between the value
of the radius of TrES-1 and that of HD~209458b, $R_p = 1.35 \pm 0.06 \ R_{\rm Jup}$,
despite the similarity in mass and degree of stellar insolation,
is an interesting puzzle with several possible resolutions.
As discussed in the following section,
\emph{HST} may play a central role in identifying the correct answer.

It is critical to note that TrES-1 is only the second transiting
planet for which the bulk of \emph{HST}-based studies (described in
the next section) may be pursued; planets identified by the
OGLE survey orbit stars that are simply too faint.  In the
following section, we review these \emph{HST}-based investigations.

\section{\emph{HST} Studies of Transiting Extrasolar Gas Giants}
Nearly the entirety of \emph{HST}-based studies of transiting exoplanets
to date has focused on HD~209458 (one exception is the ACS-HRC 
campaign to observe two transits of 
OGLE-TR-56)\footnote{\texttt{http://www.stsci.edu/cgi-bin/get-proposal-info?9805}}.  
I summarize these efforts below; for a more detailed description, see Charbonneau (2003b; 2004).

\subsection{Improved Estimates of Planetary Parameters}
Ground-based efforts to obtain precise time series photometry of
HD~209458 are limited to a precision of $\sim$2~mmag and a cadence
of $\sim$10 minutes (Charbonneau et al. 2000; Henry et al. 2000; Jha et al. 2000;
Deeg, Garrido, \& Claret 2001).  With data of this quality, 
estimations of the planetary radius (and other parameters) are
confounded by a significant degeneracy between the planetary
and stellar radii and the orbital inclination.  In short, by
increasing the planetary and stellar radii in tandem to preserve their ratio, 
the same transit depth can
be produced, and reducing the orbital inclination preserves
the chord length across the star to match the observed transit
duration.  Since independent estimates of the stellar radius from stellar 
models (e.g. Cody \& Sasselov 2002) were typically limited to a precision of 
$\sim 10$\%, estimates of the planetary radius retained this significant uncertainty.

The ultra-precise light curve by Brown et al. (2001) was produced
by using STIS to disperse the large number of photons over
as many pixels as possible (to retain a high observing efficiency
and mitigate flat-fielding effects),
and subsequently summing the recorded counts to produce a
photometric index with a typical precision of $1.1 \times 10^{-4}$ and
a cadence of 80~s.  The quality of these data breaks the
former degeneracy (principally as a result of the precise
measurement of the slope and duration of ingress and egress).
As a result, the authors derived a precise estimate of
the planetary radius, $R_{p} = 1.35 \pm 0.06 \ R_{\rm Jup}$,
as well as that of the star, $R_{s} = 1.15 \pm 0.05 \ R_{\rm Sun}$ (consistent
with, but more precise than the \emph{Hipparcos} value).
This estimate is still subject to the uncertainty in the independent
estimate of the stellar mass, $M_{s}$, but the effect is small since the
uncertainty in the radius is only weekly dependent upon that
of the mass, i.e. $\Delta R_{p} / R_{p} \simeq 0.3 \, \Delta M_{s} / M_{s}$.

More recently, Charbonneau et al. (2003c) gathered new STIS data
at lower resolution but spanning a large wavelength range, $290 - 1080$~nm.
The data can be subdivided into various effective bandpasses
prior to forming the photometric time series, and the resulting
light curves clearly show the predicted color-dependence due
to limb-darkening.  As a result, it should be possible to 
assume limb-darkening coefficients based on a stellar model,
and derive a very precise estimate of the planetary radius.

\subsection{Searches for Planetary Satellites, Circumplanetary Rings, and Reflected Light}
Brown et al. (2001) also noted that the STIS data described above
was sufficiently precise that terrestrial-sized objects present
in the HD~209458 system could be revealed.  This sensitivity was
unprecedented: Although ground-based searches for Earth-sized
objects are possible for small stars such as the M-dwarf binary
CM~Dra (e.g. Doyle et al. 2000; Deeg et al. 2000), no instrument
had previously demonstrated the ability to see such small
objects in transit across a star the size of the Sun.  Brown et al. (2001)
concluded that they could exclude the presence of a planetary satellite
with radius larger than $1.2~R_{\rm Earth}$ (although not for
all possible values of the putative satellite's orbital period and phase).  
Similarly, they excluded the presence of opaque circumplanetary rings with
a radius greater than $1.8~R_{p}$, since such a ring system
would have resulted in large deviations during ingress
and egress, which were not seen.

Brown et al. (2001) also examined their data for offsets
in the times of the center of the planetary eclipses; such offsets
would result from a massive planetary satellite.
Since the centroid of each eclipse could be measured with a precision
of 6~s, they were able to exclude satellites more massive
than $3~M_{\rm Earth}$.  More recently, Schultz et al. (2003)
used \emph{HST}'s Fine Guidance Sensors (FGS) to obtain photometry
with an exceptional cadence of 0.025~s and a
SNR of 80.  By targeting times of ingress and egress, they
will place stringent constraints on any timing variations
in this system.

Exquisite photometry of a hot-Jupiter system
might also enable the detection of the light reflected from the planet.
This effect has been sought using a spectroscopic technique from the ground for
several hot-Jupiter systems (Charbonneau et al. 1999;
Collier Cameron et al. 2002; Leigh et al. 2003), but,
as of yet, only upper limits have been obtained.  The eclipsing geometry
of the HD~209458 system presents an attractive opportunity;
precise photometry during times of secondary eclipse
could detect the decrement in light as the planet passes
behind the star.  Not only would such data enable an evaluation
of the wavelength-dependent albedo, but these data would also
constrain the orbital eccentricity through the timing of the
secondary eclipse (Charbonneau 2003b).
This project is also being pursued by
the \emph{MOST} satellite
(Walker et al. 2003; Matthews et al. 2004)\footnote{\texttt{http://www.astro.ubc.ca/MOST/}},
which has achieved unprecedented precision despite its small
aperture.  However, \emph{HST} retains the spectroscopic information
(either through STIS, or grism modes of ACS), whereas \emph{MOST}
has a single fixed bandpass.

\subsection{Atmospheric Transmission Spectroscopy}
Based on theoretical predictions (Seager \& Sasselov 2000; Brown 2001;
Hubbard et al. 2001) that hot Jupiters such as HD~209458b should
present strong alkali metal features in their transmission spectra,
Charbonneau et al. (2002) pursued this effect with STIS.  They
detected an increase in the transit depth of $(2.32 \, \pm \, 0.57) \, \times \, 10^{-4}$
in a 1.2~nm bandpass centered on the Na~D lines at 589.3~nm.  After
ruling out alternate explanations for this effect, they concluded that
this decrement was indeed due to absorption from atomic sodium
in the planetary atmosphere.  Notably, the observed signal
was only 1/3 that predicted from their fiducial model, a cloudless
atmosphere with a solar abundance of sodium in atomic form.
One explanation for this decrease may be the presence of clouds
high in the atmosphere, effectively reducing the size of
the atmosphere viewed in transmission.  Such clouds would
likely reduce other transmission spectral features in a similar
fashion, and indeed recent ground-based work in the infrared
seems to confirm this interpretation (Deming et al. 2004).

Using STIS in the UV, Vidal-Madjar et al. (2003) detected
a 15\% decrement in the flux at Ly~$\alpha$ during times of
eclipse, which they interpret as evidence for ongoing atmospheric
escape of a significant quantity of hydrogen (Lecavelier des Etangs et al. 2004).
In a more recent paper, Vidal-Madjar et al. (2004) also
present evidence of a corresponding decrement in lines
of carbon and oxygen, indicating their presence in the
cloud of material surrounding the planet; however, the effect for these
features is detected with significantly less statistical significance
than that of Ly~$\alpha$.  The precise mechanism of escape,
and the connection (if any) of this extended material to the inflated value
of the radius remain open questions that demand further study.  

\section{Connection to the \emph{Kepler} Mission}
The NASA \emph{Kepler}\footnote{\texttt{http://www.kepler.arc.nasa.gov/}}
mission (Borucki et al. 2003a; 2003b) is scheduled for launch in 2007, and may make the first
detection of an extrasolar Earth-like planet.  Indeed, if
most Sun-like stars have such planets, dozens should
be detected.  \emph{Kepler} will therefore set the scale for future efforts
to characterize directly such extrasolar Earths, since it
will tell us how common these objects are, and, by inference,
how far we might expect to look to locate the closest example.

As with the ground-based searches, \emph{Kepler} will need robust
techniques by which to discriminate true planetary transits
from astrophysical false positives, notably stellar blends.
It should be noted, therefore, that \emph{HST}/ACS is likely the only
instrument in existence that will be capable of confirming
the photometric signals detected by the \emph{Kepler} Mission.  More
importantly, \emph{HST}/ACS would be able to provide photometry
over several band passes that are distinct from the single, fixed
bandpass used by \emph{Kepler}.  As discussed in Section 3, such
photometry during times of eclipse can be used to search for 
a color-dependence to the transit depth, which would indicate
that such a candidate is an impostor.  

The following SNR calculation is adapted from one for
upcoming DD observations of TrES-1 
(T.~M. Brown, R.~L. Gilliland, et al., personal communication).  
Using ACS/HRC with the G800L grism 
to disperse the light from a $V=12.0$ star over roughly 200 pixels allows
collection of $5 \times 10^{7}$~e- per exposure while staying
30\% under saturation.  Given the much higher efficiency of
the G800L on ACS than similar capabilities on STIS, and the
significantly fainter typical magnitude for the \emph{Kepler}
candidate assumed here ($V=12.0$) than HD~209458 ($V=7.6$), it makes 
sense to use ACS (as opposed to STIS), regardless of STIS loss.
Adopting an exposure time of 115~s, and using the 512 square 
subarray nearest the readout amp for HRC (which requires a 
readout overhead of 35~s) results in a net cadence of 150~s and an
observing efficiency of 77\%.  The SNR per exposure will
be 7150.  Over the 13~hour duration of the eclipse (and accounting
for the roughly 55~minute visibility per \emph{HST}-orbit for the \emph{Kepler} f.o.v.), 
roughly 179 exposures could be obtained, for a photon-noise-limited
precision of $1.05 \times 10^{-5}$.  Since the determination of
the eclipse depth is also affected by the precision of the out-of-eclipse data,
this reduces to $1.48 \times 10^{-5}$ (for a comparable duration
out of eclipse).  This represents a 5.7-$\sigma$ detection of a Earth-sized object across a star
with the solar radius.  Of course, this calculation has not
considered the effect of systematic errors, which may limit
the precision to a level significantly worse than the theoretical
photon-noise limit.  Furthermore, \emph{Kepler} candidates that are fainter
than $V=12.0$ would be detected with less confidence,
unless the ratio of planet and stellar radii is increased.

Short-period Earth-sized planets may be detected by \emph{Kepler}
early in the mission, and \emph{HST}/ACS may still be operational
at this point.  However, true Earth analogs will be identified
only as early as 2010, since these present transits only once a year, 
and two such events are required for a period estimate (\emph{Kepler}
requires 3 events for a reliable detection).  Without a servicing mission,
it is very unlikely that \emph{HST} would still be in operation
at that time.  Given that the search for Earth-like planets is so central to 
future NASA plans, the impact of the loss of \emph{HST} to these plans
should be considered in close detail.

\section{\emph{Hubble} as a Survey Instrument}
The discussion above has focused on the use
of \emph{HST} to conduct follow-up studies of transiting extrasolar planets
identified by other telescopes.  However, \emph{HST} has also proven to be 
a powerful transit survey instrument in its own right.  The small
field-of-view precludes the simultaneous survey of
a sufficiently large number of bright stars,
as is done for many of the ground based surveys.  However,
for certain fields-of-view, \emph{HST}'s sharp point-spread function
enables precise photometry of thousands of stars for which
ground-based work cannot proceed.  To date, 
\emph{HST} has surveyed two such fields for transiting planets:
the globular cluster 47~Tuc, and the Galactic Bulge.

\subsection{A Transit Search in the Globular Cluster 47~Tuc}
The use of \emph{HST} as a survey instrument for transiting planets was
pioneered by Gilliliand et al. (2000).
In July 1999, they monitored 34,000 main-sequence 
stars in the globular cluster 47~Tuc for 8.3 days.  Star clusters
make attractive hunting grounds for transiting planets
for several reasons.  First, the uniformity of age
and metallicity of the cluster members provides the
ideal laboratory setting within which to study the dependence of 
the planet population on these quantities.  Second, since the stellar 
radii may be inferred from measurements of the
apparent brightness and an evolutionary models for the cluster,
the radius of the transiting secondary can also be
reliably inferred.  A key element in the design of
the Gilliland et al. (2000) experiment was that a large
range of stellar radii could be monitored for transiting
planets, since the poorer photometric precision obtained
for much smaller (and less massive stars) was counterbalanced
by a corresponding increase in the depth of the transit.
In particular, the stars surveyed ranged in size from 
roughly $1.5~R_{\rm Sun}$ (corresponding to slightly evolved
stars with masses of $0.87~M_{\rm Sun}$, at visual magnitudes
of $V=17.4$) to main-sequence stars with radii of $0.51~R_{\rm Sun}$
(corresponding to masses of $0.55~M_{\rm Sun}$ and a visual magnitudes 
of $V=21.9$).

The survey found no stars presenting planetary transits,
which was very much at odds with the findings of the Doppler
surveys:  In the local solar neighborhood, roughly 1\% of Sun-like
stars have hot Jupiters; folding in the recovery rates
and efficiencies specific to the \emph{HST} 47~Tuc survey, approximately
17 such objects would have been expected.  The core result
of the survey was thus that the population of hot Jupiters
in 47~Tuc was at least an order of magnitude below that of
the local solar neighborhood.  This disparity may be 
due in part to the decreased metallicity of 47~Tuc (${\rm [Fe/H]}=-0.7$),
as the lower-metallicity environment may impede planet
formation and/or migration.  A likely additional effect
is that of crowding:  At the typical location of the
Gilliland et al. (2000) observations (1~arcminute from the
core), the stellar density is roughly $10^3 \, M_{\rm Sun} \, {\rm pc}^{-3}$;
gas giant planets at several AU from the star 
could be disrupted in the typical planetary system if
it suffers a close dynamical encounter prior to the inward 
migration of the planet (at which point it is sufficiently bound to withstand
such disruption).

Due to the uniformity of the stellar population and the 
data set, and the extensive set of detection tests 
performed, the Gilliland et al. (2000) result remains
one of the few well-characterized (and hence astrophysically
useful) null results for a survey of transiting extrasolar
planets.  Wide-field surveys such as TrES are of
great interest because they may deliver bright, transiting
planets for which the parameters can be accurately measured, and which
are amenable to follow-up studies.  However, these wide-field
surveys are unlikely to yield meaningful constraints of the rate-of-occurrence
of hot Jupiters as a function of the properties and environment
of the central star, due to the very diverse nature of the targets 
surveyed in their magnitude-limited field samples.

\subsection{A Transit Search in a Galactic Bulge Field}
K. Sahu and collaborators 
(program GO-9750\footnote{\texttt{http://www.stsci.edu/cgi-bin/get-proposal-info?9750}})
have recently undertaken an ambitious \emph{HST}-based transit
search that seeks to capitalize on the increased (relative to WFPC2) 
field-of-view and sensitivity allowed by ACS/WFC.  In February 2004, 
they observed a field in the galactic bulge for 7~days with ACS/WFC.  
In the $202 \times 202$-arcsecond field-of-view, they expect to monitor
roughly 167,000 F, G, and K dwarfs brighter than $V=23$, for
which they will obtain a photometric precision sufficient to detect 
the transit of a Jupiter-sized planet.  If the rate of occurrence
of hot Jupiters for these stars is the same as that in the
solar neighborhood (roughly 1\%), they may detect more
than 100 such planets.  The number of disk and
bulge stars are approximately equal, and the membership of
a given star to a population will be determined by proper
motions from data obtained at a different epoch (this
data is also in hand).  Furthermore, the metallicity of stars in
the field-of-view is expected to vary by more than 1.5 order
of magnitude.  As a result, this dataset might permit
the detection of a sufficiently large and diverse group of hot Jupiters 
so that the dependence of rate-of-occurrence and planetary radius 
upon several properties of the primary (notably stellar type, disk 
vs. bulge membership, and metallicity) could be disentangled.

A note of caution must be sounded, however.  As discussed earlier,
the primary challenge facing transit surveys is no longer that
of photometric precision and phase coverage, but rather
that of rejecting the astrophysical false positives, whose
photometric light curves precisely mimic that of a 
planetary transit (Charbonneau et al. 2004, Latham 2003).  
For the brightest targets, VLT radial velocity monitoring may 
reveal the spectroscopic orbit.  More importantly, a detailed 
spectroscopic analysis for blends of eclipsing binaries could 
be performed.  However, even for much brighter stars that have been surveyed
for transits it has proven extremely difficult to
identify hierarchical triples, in which the light from a third, 
bright star dilutes the photometric and spectroscopic variability of the
eclipsing binary.  Two examples of the degree to which 
such systems may confound researchers are given by
Torres et al. 2004b (for a system identified by the OGLE survey,
Udalski et al. 2002a, 2002b, 2003) and Mandushev et al. 2004 (for
a candidate found by the TrES network).  For the bulk of candidates, 
however, no spectrograph in existence will be able to recover the
Doppler orbit (for a detailed presentation of the signal-to-noise
calculation, see Charbonneau 2003a).  The investigators will then need to
rely upon more indirect considerations (such as colors and proper motions)
to argue in favor of a planetary interpretation.  Whether
this argument can be made convincingly for an object
that displays a transit light curve but no measurable
Doppler orbit remains to be seen.  Moreover, since the planetary
radius is approximately degenerate with mass across two orders
of magnitude ($0.5-80~M_{\rm Jup}$; Burrows et al. 2001), 
the value of each individual object will
be diminished relative to those identified for brighter stars.
These concerns aside, however, the prospect of perhaps
doubling the number of known extrasolar planetary systems
is a fascinating one indeed, and the results from this survey
are eagerly awaited.

\section{Final Note Regarding STIS}
As I was finalizing this contribution, the STIS spectrograph 
went offline due to the failure of an internal 5V power
supply\footnote{\texttt{http://www.stsci.edu/hst/stis/}}.  As described above, 
STIS was the most productive \emph{HST} instrument with regards to
follow-up studies of hot Jupiters.  Fortunately, many investigations
requiring high photometric and/or spectroscopic stability can be 
accomplished with ACS (using a grism element to handle to the high 
photon rates), as well as NICMOS and FGS.  Notably, the typical brightness
of the transiting hot Jupiters that will be identified over the
next couple years by the numerous ongoing wide-field surveys is optimally
suited for ACS rather than STIS, regardless of STIS loss.   

\end{document}